\documentclass[aps,prl,floatfix,nofootinbib,showpacs,twocolumn,amsmath,amssymb]{revtex4}
\usepackage{dcolumn}
\usepackage{bm}
\usepackage[dvips]{epsfig}
\usepackage{graphicx} \usepackage{amsmath} \usepackage{amssymb}
\newcommand{\BEQ}{\begin{equation}}
\newcommand{\EEQ}{\end{equation}}
\newcommand{\BEA}{\begin{eqnarray}}
\newcommand{\EEA}{\end{eqnarray}}

\renewcommand{\d}{{\rm d}}

\newcommand{\e}{\epsilon}

\newcommand{\om}{\omega}

\newcommand{\ba}{\bar{a}}

\newcommand{\bp}{\bar{p}}
\renewcommand{\ba}{\bar{a}}

\begin{document}

\title
{Replicators in Fine-grained Environment: Adaptation and
Polymorphism\footnote{Published in Phys. Rev. Lett. {\bf 102}, 058102 (2009)}
}
\date{\today}

\author{Armen E. Allahverdyan$^{1)}$ and Chin-Kun Hu$^{2,3)}$}

\affiliation{$^{1)}$Yerevan Physics Institute,
Alikhanian Brothers Street 2, Yerevan 375036, Armenia\\
$^{2)}$ Institute of Physics, Academia Sinica, Nankang, Taipei
11529, Taiwan\\
$^{3)}$Center for Nonlinear and Complex Systems and Department of
Physics, Chung Yuan Christian University, Chungli 32023, Taiwan}

\begin{abstract}

Selection in a time-periodic environment is modeled via the
two-player replicator dynamics. For sufficiently fast
environmental changes, this is reduced to a multi-player
replicator dynamics in a constant environment. The two-player
terms correspond to the time-averaged payoffs, while the three and
four-player terms arise from the adaptation of the morphs to their
varying environment.
Such multi-player (adaptive) terms can induce a stable
polymorphism. The establishment of the polymorphism in partnership
games [genetic selection] is accompanied by decreasing mean
fitness of the population.

\end{abstract}

\pacs{87.23.-n, 87.23.Cc, 87.23.Kg, 02.50.Le}






\maketitle

Environmental impact on adaptation, selection and evolution is an
important subject of biological research
\cite{levins,bull,grant,cook,stein,janavar,svi,gill,clark}.
The main scenario of environmental adaptation on the population
level is polymorphism
\cite{levins,stein,bull,grant}:
two or more clearly different types of phenotype (morph) exist in
one interbreeding population. The basic mechanisms of polymorphism
are heterozygote advantage and inhomogeneous (frequency, space
and/or time-dependent) environment \cite{stein,grant}.
Polymorphism can be restricted to the phenotype level, or it may
be controlled genetically by multiple alleles at a single locus ,
e.g., human ABO blood groups \cite{grant}.
Here is one example of polymorphism related to a time-dependent
environment \cite{grant,cook}.  Forrest populations of the land
snail {\it Cepaea Nemoralis} consist of three morphs having
respectively brown, pink and yellow colored shells
\cite{grant,cook}.  The brown and pink morphs have an advantage
over the yellow morph at the spring time, since the background
color makes them less visible for predators \cite{grant,cook}; the
yellow morph has an advantage at summer and autumn on the
yellow-green substrate.  In addition, the yellow morph is more
resistant to high and low temperatures \cite{grant,cook}. Thus
different morphs have different advantage under different
environmental conditions \cite{grant}.

A varying environment has roughly three dimensions: it may be time
or/and space dependent, predictable {\it vs} stochastic, and fine
{\it vs} coarse-grained \cite{levins}. The latter means that each
individual within population sees mainly one fixed environment,
which can change from one generation to another, for example. A
fine grained environment changes many times during the life-time
of each individual; see the above example of {\it Cepaea} and note
that this snail lives seven to eight years \cite{grant,cook}.

Much attention was devoted to modeling polymorphism in various
coarse-grained environments
\cite{levins,stein,janavar,svi,gill,clark}.  Fine-grained
environments got less attention, since early theoretical results
\cite{levins,strobeck} and the biological common sense
\cite{stein} implied that non-trivial polymorphism scenarios are
absent.  One expects that in this case the organism sees (thus
adapts to) the average environment \cite{levins,stein,strobeck}.
However, recent experiments indicate that the evolving populations
can adapt to time-varying aspects of their fine-grained environment
\cite{miner_vonesh,winn,jasmin}.  In particular, they can respond
to the environmental patterns other than the environmental mean
\cite{miner_vonesh}.  Moreover, the total fitness during such an
adaptation need not increase \cite{jasmin}. A proper theoretical
model for such phenomena is still absent.

Here we present a theory for polymorphism in fine-grained,
time-periodic environment based on Evolutionary Game Theory (EGT).
Our main method is the time-scale separation in the replicator
dynamics.

EGT describes interacting agents separated into several groups
\cite{svi,hofbauer}. The reproduction of each group is governed by
its fitness, which depends on interactions between the groups. The
most popular replicator dynamics approach to EGT describes the
time-dependent frequency $p_k(t)$ of the group $k$, which is the
number of agents $N_k$ in the group $k$, over the total number of
agents in all $n$ groups: $p_k ={N_k}/{\sum_{k=1}^n N_k}$. The
fitness $f_k$ of the group $k$ is a linear function of the
frequencies \cite{svi,hofbauer}:

\BEA \label{z1} f_k(a,p)={\sum}_{l=1}^n
a_{kl} p_l, \qquad k=1,\ldots, n,
\EEA
where the {\it payoffs}
$a_{kl}$ account for the interaction between (the agents from)
groups $k$ and $l$. The replicator dynamics \cite{hofbauer}
facilitates the (relative) growth of groups with fitness larger
than the average fitness ${\sum}_{l=1}^n p_l f_l$: \BEA \label{z2}
\dot{p}_k=p_k [\,f_k(a,p)-{\sum}_{l=1}^n p_l f_l\,]\equiv
G_k[a,p]. \EEA Within game theory the groups correspond to
strategies, while the pay-offs $a_{kl}$ describe interaction
between two players: the probability $p_k$ of strategy $k$ changes
according to the average pay-off $\sum_l a_{kl}p_l$ received by
one player in response to applying the strategy $k$
\cite{svi,hofbauer}.

There are several applications of EGT and replicator dynamics in
biology: {\it i)} Animal (agent) contests, where the groups
correspond to the strategies of agent's behavior during the
contest, while $p_k$ is the probability by which an agent applies
the strategy $k$ \cite{hofbauer}. The actual mechanism by which
$p_k$ changes depends on the concrete implementation of the model
(inheritance, learning, imitation, infection, etc).  {\it ii)}
Selection of genes, where $p_k$ is the frequency of one-locus
allele $k$ in panmictic, asexual, diploid population, and where
$a_{kl}=a_{lk}$ refers to the selective value of the phenotype
driven by the zygote $(kl)$ \cite{svi}.  Then (\ref{z1}, \ref{z2})
are the Fisher equations for the selection with overlapping
generations \cite{svi,hofbauer}. {\it iii)} The basic
Lotka-Volterra equations of ecological dynamics can be recast in
the form (\ref{z2}) and studied as replicators \cite{hofbauer}.
Within the replicator approach polymorphism means a stable state,
where two or more $p_k$ are non-zero.


We consider a varying, but predictable environment, which acts on
the phenotypes making $a_{kl}$ periodic functions of time with
period $2\pi/\om$ \cite{svi,broom}: $ a_{kl}(\tau
)=a_{kl}(\tau+2\pi), ~~ \tau\equiv\om t$. There are well-defined
methods to decide to which extent a varying environment is
predictable for a given organism \cite{clark}.  The oscillating
payoffs can reflect the fact that different morphs (alleles,
strategies) are dominating at different times. Let us assume that
the environment varies fast [fine-graining]: the average change of
the population structure over the environment period
${2\pi}/{\om}$ is small. We separate the time-dependent payoffs
$a_{kl}(\tau )$ into the constant part $\bar{a}_{kl}$ and the
oscillating part $\widetilde{a}_{kl}(\tau)$ with zero time-average
$\bar{\widetilde{a}}_{kl}\equiv \int_0^{2\pi}\frac{\d
\tau}{2\pi}\, \widetilde{a}_{kl}(\tau)=0$, \BEA \label{6}
a_{kl}(\tau)=\bar{a}_{kl}+\widetilde{a}_{kl}(\tau),\quad
\partial_\tau\hat{a}_{kl}(\tau)=\widetilde{a}_{kl}(\tau), \quad \overline{\hat{a}_{kl}}=0,
\EEA where $\hat{a}_{kl}$ is defined to be the primitive of
$\widetilde{a}_{kl}$ with its time-average equal to zero.

Following the Kapitza method \cite{ll}, we represent $p_k$ as a
slowly varying part $\bp_k$ plus $\e_k[\bp (t),\tau]$, which is
smaller than $\bp_k$, fast oscillating on the environment time
$\tau$, and averaging to zero: \BEA \label{8}
&&p_k(t)=\bp_k(t)+\e_k[\bp (t), \tau],\\
\label{8.1} &&\bar{\e}_k[\bp (t)]=\int_0^{2\pi}\frac{\d
\tau}{2\pi}\, \e_{k}[\bp (t),\tau]=0. \EEA Here the average is
taken over the fast time $\tau$ for a fixed slow time $t$. Note
that the fast $\e_k$ depends on the slow $\bp$. Now put (\ref{8})
into (\ref{z2}) and expand the RHS of (\ref{z2}) over $\e$: \BEA
\dot{\bp}_k+\dot{\bp}_\alpha\partial_\alpha \e_k+\om \partial_\tau
\e_k =[1+\e_\alpha\partial_\alpha +{\cal
O}(\e^2)]G_k[\widetilde{a}(\tau),\bp], \label{z5} \EEA where the
summation over the repeated Greek indices, is assumed and
$\partial_\alpha X\equiv \frac{\partial }{\partial \bp_\alpha}X$.
The fast factor $\e_i$ is searched for via expanding over
$\frac{1}{\omega}$: $\e_k= \frac{1}{\omega}
\e_{k,1}+\frac{1}{\omega^2} \e_{k,2}+\ldots$. Substitute this into
(\ref{z5}), add and subtract suitable averages, and get for the
fast terms: $\partial_\tau
\e_{k,1}=G_k[\widetilde{a}(\tau),\bp]+{\cal O}(\frac{1}{\omega})$.
After straightforward integration, we have \BEA \label{z8}
\e_k[\bp(t), \tau]= \frac{1}{\om}\, G_k[\hat{a}(\tau),\bar{p}(t)]
+{\cal O}(\frac{1}{\omega^2}), \EEA where
$\hat{a}=\{\hat{a}_{kl}\}$ is defined in (\ref{6}). Once
(\ref{z8}) is separated, the remainder in (\ref{z5}) is the
evolution of slow terms
\begin{gather}
\label{9}
\dot{\bp}_k=G_k[\bar{a}, \bp]
+\overline{\, \e_\alpha[\bp,\tau] \,\,\partial_\alpha
G_k[\widetilde{a}(\tau), \bp]  \, }+{\cal O}({1}/{\omega^2}),
\end{gather}
where the time-average is defined as in (\ref{8.1}).
Working out (\ref{9}) and using
$\overline{\hat{a}\widetilde{b}}=-\overline{\widetilde{a}\hat{b}}$
we get again a replicator equation
\BEA
\label{11}
\dot{\bp}_k
&=&\bp_k\left[{\cal F}_k(\bp)-\bp_\alpha {\cal F}_\alpha(\bp)\right], \\
\label{12}
{\cal F}_k(\bp) &=&
\bar{a}_{k\alpha} \bp_\alpha
+b_{k\alpha\beta} \bp_\alpha\bp_\beta
+c_{k\alpha\beta\gamma} \bp_\alpha\bp_\beta\bp_\gamma
\EEA
where ${\cal F}_k$ is the effective (already non-linear) fitness, and
\BEA
\label{14}
b_{klm}\equiv\frac{1}{\om}\,
\overline{\widetilde{a}_{kl}\,\hat{a}_{lm}},
\qquad
c_{klmn}\equiv \frac{1}{\om}\,
\overline{\hat{a}_{kl} \, \widetilde{a}_{mn}^{[{\rm s}]}  }.
\EEA Here we defined $a_{kl}^{[{\rm
s}]}\equiv\frac{1}{2}(a_{kl}+a_{lk})$. Equations~(\ref{11},
\ref{12}) are our central result. Expectedly, the fast environment
contributes the averaged payoffs $\bar{a}_{kl}$ into ${\cal F}_k$.
This is well-known for any fine-grained environment \cite{levins}.

\begin{figure}
\includegraphics[width=4cm]{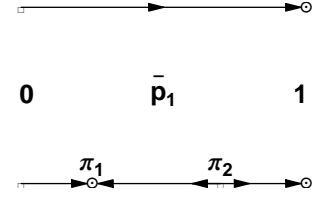}
\caption{ Schematic portrait of (\ref{21}) for $C=0$ [upper diagram] and for $C$
satisfying conditions (\ref{24}) [lower diagram]. Stable [unstable] fixed points are denoted
by circles [squares]. Arrows indicate direction of flow in time.
}
\label{f_0}
\end{figure}

However, besides this averaged two-party interaction, each group
$k$ gets engaged into three- and four-party interactions with
payoffs $b_{klm}$ and $c_{klmn}$, respectively. Indeed, recalling
our discussion after (\ref{z2}), we can interpret
$\sum_{lm}b_{klm}\bp_l\bp_m$ in (\ref{12}) as the average pay-off
received by one of three players upon applying strategy $k$.
The terms with $b_{klm}$ and
$c_{klmn}$ in (\ref{12}) exist due to adaptation of the morphs to
their environment: while the frequencies of the morphs fast
oscillate on the environmental time, see (\ref{8.1}), on longer
times the population sees the fast environment as an effective
many-player model.  Note that the terms with $b_{klm}$ and
$c_{klmn}$ need not be small as compared to $\bar{a}_{kl}$-terms,
since the derivation of (\ref{11}--\ref{14}) applies for $\bp_k\gg
\e_k$, which can hold even for $\bar{a}_{kl}\to 0$.

We get $b_{klm}=c_{klmn}=0$ [due to
$\overline{\hat{a}_{kl}\widetilde{a}_{kl}}=0$], if only one
$a_{kl}$ varies in time, or if all $\widetilde{a}_{kl}$ oscillate
at one phase: $\widetilde{a}_{kl}=a(t)\xi_{kl}$, where $\xi_{kl}$
are constant amplitudes.  Indeed, the adaptive terms [with
$b_{klm}$ and $c_{klmn}$] are non-zero due to interference between
the environmental oscillations of $\widetilde{a}_{kl}$ and those
of $\e_k$, which are delayed over the environmental oscillations
by phase $\pi/2$; see (\ref{z8}). This is why there is no
interference if all $\widetilde{a}_{kl}$ oscillate in phase.  Thus
for having the adaptive terms we need at least two morphs reacting
on the environment {\it differently}, e.g., due to different delay
times of reaction.  For the adaptive terms we also need the
frequency-dependent selection, e.g., they are absent for haploid
replicators $\dot{p}_k=p_k[a_k(t)-a_\alpha(t) p_\alpha]$.


Euation~(\ref{11}) implies that the relative growth of two morphs
at slow (long) times is determined by the effective fitness
difference: $\frac{\d}{\d t}(\bp_k/\bp_l)=(\bp_k/\bp_l)({\cal
F}_k-{\cal F}_l)$. Thus in the stable fixed points of (\ref{11})
the (effective) fitness of surviving morphs are equal to each
other, while the fitness of non-surviving ($\bar{p}_k\to 0$)
morphs is smaller (Nash equilibrium) \cite{hofbauer}. Another
pertinent quantity, the (fast) time-averaged fitness
$\bar{f}_k[p(t)]$ is the cumulative effect of the short-time
replication intensities.  Employing (\ref{8}, \ref{8.1}, \ref{z8})
we deduce that {\it even when} the adaptive terms are taken into
account, these two quantities are equal ${\cal
F}_k=\overline{f_k(\bar{a}+\widetilde{a}, \bp+\e)}$. The overall
long-time fitness of the population is characterized by the
effective mean fitness $\bp_\alpha {\cal F}_\alpha$, which is also
equal to its time-averaged analog: \BEA \Phi
&\equiv&\overline{(\bp_\alpha+\e_\alpha)(\bar{a}_{\alpha\beta}+\widetilde{a}_{\alpha\beta})
(\bp_\beta+\e_\beta)}=\bp_\alpha {\cal F}_\alpha\nonumber \\
&=&\bar{a}_{\alpha\beta}\bp_\alpha\bp_\beta
+b_{\alpha\beta\gamma}\bp_\alpha\bp_\beta\bp_\gamma.
\label{kondor}
\EEA
The contribution $c_{\alpha\beta\gamma\delta}\bp_\alpha\bp_\beta\bp_\gamma\bp_\delta$
in $\bp_\alpha {\cal F}_\alpha$ nullifies due to (\ref{14}) and
$\overline{\hat{a}\widetilde{b}}=-\overline{\widetilde{a}\hat{b}}$.

The mean fitness $\Phi$ is especially important for partnership
games [genetic selection]: $a_{ik}(\tau)=a_{ki}(\tau)$, since for
the constant payoff situation in the replicator equation
(\ref{z2}), $\Phi$ monotonically increases  towards its nearest
local maximum over the set of variables $\bp_k$ (fundamental
theorem of natural selection) \cite{svi,hofbauer}.  As follows
from (\ref{14}, \ref{kondor}), for $a_{ik}(\tau)=a_{ki}(\tau)$,
$\Phi$ reduces to the averaged two-player contribution:
$\Phi=\sum_{\alpha\beta}\bp_\alpha\,\bar{a}_{\alpha\beta}\,\bp_\beta$.
However, the theorem is not valid in the presence of the adaptive
terms, and the mean fitness can decrease; see below.

Let us now study concrete examples.
For $n=2$, Eq.~(\ref{z2}) simplifies to
a closed equation for the frequency $p_1$
\begin{gather}
\label{18}
\dot{p}_1=p_1(1-p_1)[A(t)-B(t)p_1],\\
A(t)\equiv a_{12}(t)-a_{22}(t),
B(t)\equiv 2a_{12}(t)-a_{11}(t)-a_{22}(t),\nonumber
\end{gather}
where without much loss of generality we adopted $a_{12}(\tau)=a_{21}(\tau)$.
Using notations (\ref{6}) and defining
\BEA
\label{u1}
C\equiv\frac{1}{\om}\, \overline{\hat{A}\,\widetilde{B}}
=\frac{1}{\om}[\, \overline{\hat{a}_{12}\,(\widetilde{a}_{22}-\widetilde{a}_{11})}
+\overline{\hat{a}_{22}\,\widetilde{a}_{11}}\,],
\EEA
for the adaptive factor, we deduce from (\ref{z8}, \ref{9}, \ref{18})
\begin{gather}
\label{21}
\dot{\bp}_1=\bp_1(1-\bp_1)[\bar{A}-\bar{B}\bp_1-C\bp_1(1-\bp_1)].
\end{gather}
$C$ vanishes for the symmetric homozygotes,
${a}_{11}(\tau)={a}_{22}(\tau)$, and for one recessive allele,
e.g. ${a}_{12}(\tau)={a}_{22}(\tau)$. Both cases are easily
solvable from (\ref{18}) showing that the long-time behavior of
$p_1$ is indeed governed by $\bar{A}$.

The vertices $\bp_1=1$ and $\bp_1=0$ are always fixed points of (\ref{21}), while
two interior fixed points are
\BEA
\pi_{1,2}=\frac{1}{2C}[\bar{B}+C\mp \sqrt{(\bar{B}+C)^2-4\bar{A}C}\,\,],\,\,\, \pi_1<\pi_2.
\EEA
If $\pi_1$ and $\pi_2$ are in $(0,1)$, then
$\pi_1$ ($\pi_2$) is stable (unstable). The analysis of (\ref{21}) reduces to the following
scenarios.

{\bf 1.} For $\bar{A}>0$ and $\bar{A}>\bar{B}$ [i.e.,
$\bar{a}_{11}>\bar{a}_{12}>\bar{a}_{22}$] the morph 1 globally
dominates for $C=0$, i.e., for all initial conditions $\bp_1$ goes
to $1$ for large times; see Fig.~\ref{f_0}. The global dominance
does not change for $C<0$. One can call this morph generalist
\cite{stein}, since its fitness does not oscillate in time ($\e_k$
in (\ref{z8}) is zero for $\bp_1=1$), and its fitness is maximal;
see also below. For \BEA \label{24} C>\bar{B}>0~~{\rm and}~~
(\bar{B}+C)^2\geq 4\bar{A}C, \EEA both $\pi_1$ and $\pi_2$ fall
into the interval $(0,1)$, see Fig.~\ref{f_0}, while if one of
conditions (\ref{24}) does not hold, both $\pi_1$ and $\pi_2$ are
not in this interval. Thus if (\ref{24}) hold, a stable fixed
point $\pi_1$ emerges, which attracts all the trajectories that
start  from $\bp_1(0)<\pi_2$: the polymorphism is created by the
adaptation term $\propto C$ in (\ref{21}). Initial condition
larger than the unstable fixed point $\pi_2$, $\bp_1(0)>\pi_2$,
still tend to $\bp_1=1$; see Fig.~\ref{f_0}. Both stable fixed
points $\pi_1$ and $1$ are Evolutionary Stable States (ESS),
meaning that they cannot be invaded by a sufficiently small mutant
population \cite{hofbauer}. The coexistence of two ESS one of
which is interior (i.e., polymorphic) is impossible for a
two-player replicator equation with constant pay-offs
\cite{hofbauer}, but it is possible for multi-player replicator
equation \cite{broom_vick}. We thus saw above an example of this
behavior induced by time-varying environment.

\begin{figure}
\includegraphics[width=5.5cm]{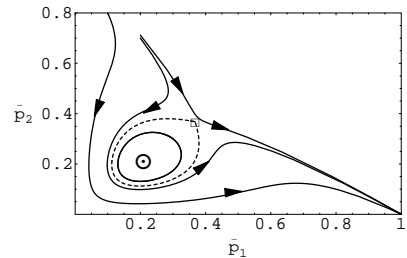}
\caption{ The portrait of (\ref{31}) for $\bar{a}_{12} =
0.1$, $\bar{a}_{13} = 0.1$, $\bar{a}_{23} = -0.1$, $\kappa_1 = -0.65$ and
$\kappa_2 = 0.65$; $\bar{p}_1$ and $\bar{p}_2$ are restricted by $0\leq
\bar{p}_1\leq 1$, $0\leq \bar{p}_2\leq 1$, $\bar{p}_1+\bar{p}_2\leq 1$. Two fixed points
are denoted by square (saddle) and cycle (center).
The closed orbits contain the center in
their interior; orbits from the second class converge to $\bar{p}_1=1$. These
two classes of orbits are separated by a dashed curve (separatrix), which is made by joining
together two unstable directions of the saddle. Arrows indicate direction of flow.
}
\label{f_1}
\end{figure}

As we discussed below (\ref{kondor}), the mean fitness does not
contain the adaptive terms directly and is given as
$\Phi=2\bar{A}\bp_1-\bar{B}\bp_1^2$ (up to an irrelevant
constant). For $\bar{A}>0$ and $\bar{A}>\bar{B}$, $\Phi$ maximizes
at $\bp_1=1$, and this maximum is the only stable fixed point of
the replicator dynamics (\ref{21}) with $C=0$. If however $C$
satisfies conditions (\ref{24}), in the stable fixed point
$\bp_1=\pi_1$ the mean fitness $\Phi$ is smaller than at the
stable point $\bp_1=1$. Moreover, for the initial conditions
$\pi_1<\bp_1(0)<\pi_2$, the mean fitness $\Phi$ decreases in the
course of the relaxation to $\pi_1$.

The quantity which is increased by dynamics (\ref{21}) is
\BEA
\Psi=2\bar{A}\bp_1-\bar{B}\bp_1^2
-C\bp_1^2(1-\frac{2\bp_1}{3})=\Phi-C\bp_1^2(1-\frac{2\bp_1}{3}).\nonumber
\EEA
Thus $\Psi$ is the Lyapunov function: $\dot{\Psi}\geq 0$.  Though $\Psi-\Phi<0$ whenever the
adaptive polymorphism conditions (\ref{24}) hold, the difference between
the Lyapunov function and the mean fitness is not negative for $C<0$.
Compare these facts to the evolution in a coarse-grained environment, which
tends to maximize the average fitness {\it minus a positive} risk aversion factor; see
\cite{aversion} for reviews.

{\bf 2.} For $\bar{B}>\bar{A}>0$ [i.e., $\bar{a}_{12}>\bar{a}_{11},
\bar{a}_{22}$] and $C=0$ there is a stable polymorphism at the fixed
point $\bp_1={\bar{A}}/{\bar{B}}$ (heterozygote advantage).  The
presence of $C\not =0$ in (\ref{21}) does not change this polymorphism;
only the value of the fixed point shifts to $\pi_1$. In contrast to the
scenario {\bf 1}, here the response to slow environmental changes is
reversible.


{\bf 3.} For $\bar{B}<\bar{A}<0$ and $C=0$ there is an unstable
polymorphism: all the initial conditions with
$\bp_1(0)<{\bar{A}}/{\bar{B}}$ end up at $p_1=0$ (morph 2 dominates),
while those with $p_1(0)>{\bar{A}}/{\bar{B}}$ finish at $p_1=1$ (morph 1
dominates).  Now $C\not =0$ in (\ref{21}) shifts the
unstable fixed point to $\pi_2$.

These are all possible scenarios for $n=2$; other relations
between $\bar{A}$ and $\bar{B}$ lead to interchanging morphs.

For three morphs, $n=3$, we assume the zero-sum
situation in (\ref{z2}), $a_{kl}(\tau)=-a_{lk}(\tau)$ \cite{akin}: the loss of the
strategy $l$ is equal to the gain of $k$.
Equations~(\ref{11}--\ref{14})  reduce to
\BEA \label{31}
\dot{\bp}_i=\bp_i \bar{a}_{i\alpha}\bp_\alpha +\kappa_i
\bp_1\bp_2\bp_3,\,\, \kappa_1=2b_{123}, \,\, \kappa_2=2b_{213},
\EEA where $\sum_{i=1}^3\kappa_i=0$. The four-party contribution
disappears from (\ref{31}).  One can show that any interior fixed
point of (\ref{31}) can be either saddle (two real eigenvalues of
the Jacobian with different sign) or center (two imaginary,
complex conjugate eigenvalues).

{\bf 4.} For $\ba_{12}>0$, $\ba_{13}>0$ and $\kappa_i=0$ the morph
1 globally dominates: $\bp_1=1$ is the only stable fixed point.
For the existence of polymorphism it is necessary that
$\kappa_1<0$, i.e., the strategies 2 and 3 {\it together} win over
1, although {\it separately} they lose to 1.  The dominance of 1
is still kept when $\bp_2$ or $\bp_3$ are forced to decay, because
then the adaptive term in (\ref{31}) is irrelevant for
sufficiently large times. This happens when $\ba_{23}<0$ and
$\kappa_2<0$ or when $\ba_{23}>0$ and $\kappa_3<0$. Apart from
these cases the terms $\propto \kappa_i$ in (\ref{31}) can lead to
polymorphism, provided that their magnitude is large enough; see
Fig.~\ref{f_1}. Besides the stable fixed point $\bp_1=1$ of the
$\kappa_i=0$ dynamics, two new fixed points emerge: stable
(center) and unstable (saddle).  A domain around the saddle
supports polymorphism with cyclic dominance of the morphs; see
Fig.~\ref{f_1}.  We see a general feature of all the above
examples: the adaptive (multi-party) terms do not influence the
local stability of the vertices (where all but one $\bp_k$'s are
zero).


{\bf 5.} For $\ba_{12}>0, \quad \ba_{13}<0, \quad \ba_{23}>0$
there is polymorphism already for $\kappa_i=0$: 1 wins over 2,
which wins over 3, but 3 wins over 1 (rock-scissor-paper game).
Now for $\kappa_i=0$ in (\ref{31}) there is already one interior
fixed point, and the trajectories are closed orbits around this
fixed point.  After including the adaptive ($\propto \kappa_k$)
terms in (\ref{31}) this fixed point is simply shifted, and no new
fixed points appear for any size or magnitude of $\kappa_i$.

{\it To summarize}, we have shown that in addition to the averaged
payoffs, a fast [fine-grained] time-periodic environment generates
adaptive, multi-player terms in the replicator dynamics, provided
that at least two morphs react on the environment differently.
These terms can create a polymorphic stable state via adaptation
of the ``weak'' morphs to environmental changes. This polymorphism
is related to decreasing mean fitness of the population. This
specific aspect of the polymorphism was argued to be a
prerequisite for the phenomenon of sympatric speciation, where by
contrast to the allopatric scenario the speciation is induced
inside a single population \cite{doeb}. Thus, our results hint at
a sympatric speciation scenario due to a fine-grained,
time-periodic environment.




We thank M. Broom and K. Petrosyan for discussions, and K.-t.
Leung for critical reading. The work was supported by
Volkswagenstiftung, grants NSC 96-2911-M 001-003-MY3 \&
AS-95-TP-A07, and National Center for Theoretical Sciences in
Taiwan.


\end{document}